\documentclass[letterpaper]{article}
\usepackage{amsmath}

\def\art#1{[\ref{#1}]}

\begin{document}

\title{\large{On the Comment on ``CCC-predicted low-variance circles in CMB sky and LCDM'', by H.K.~Eriksen and I.K.~Wehus}}
\author{A.A.~Kocharyan\footnote{armen.kocharyan@monash.edu}\\
{\small School of Mathematical Sciences, Monash University, Clayton, Australia}}

\maketitle

\begin{center}
{\bf Abstract}
\end{center}

In \art{EW}, Eriksen and Wehus try to criticise the use of Kolmogorov's stochasticity parameter (KSP) for CMB data analysis. Their discussion is based on a serious misunderstanding of the randomness and Arnold's work. Their note includes numerous further inaccuracies and groundless statements. However, this short note is concerned with KSP method only. 

\vspace{0.3in}

How does one know if a finite sequence of real numbers is random or non-random? No doubts that nobody would agree with handwaving arguments in~\art{EW}.

Without defining randomness or providing meaningful arguments, not to say about doing alternative computations, Eriksen and Wehus write: {\it ``When reading these papers, it seems clear to us that Gurzadyan et al. confuse randomness with correlation: While the CMB field is ({\textit{\textbf{most likely}}}) a random field, it is not uncorrelated. Instead, the CMB field is a smooth field on scales comparable with the instrumental beam, and it has a well-defined non-flat power spectrum. Thus, the real-space correlations are strong. Of course, the instrumental noise is virtually uncorrelated, and so there are indeed two components here, one correlated and one uncorrelated. 
{\textit{\textbf{But neither is non-random.}}}''}
\footnote{All emphasis are ours.}

This is not how scientific questions are addressed. Arnold shows how it is properly done in \art{A_KSP}, \art{Arnold_UMN}, \art{Arnold_MMS}, \art{Arnold_FA}. The stochastic parameter and the statistic introduced by Kolmogorov \art{K} is applied to measure the objective stochasticity degree of datasets. He proves that KSP method is mathematically sound, non-trivial, and universal.

For instance, the following sequence
\begin{center}
A=\{3, 9, 27, 81, 43, 29, 87, 61, 83, 49, 47, 41, 23, 69, 7\},
\end{center}
looks as random as the sequence
\begin{center}
B=\{37, 74, 11, 48, 85, 22, 59, 96, 33, 70, 7, 44, 81, 18, 55\}.
\end{center}

Arnold shows, using KSP method, that randomness probability is higher for $A$ than for $B$ \art{A_KSP}.

The approach in \art{GK_KSP} is based on a solid ground built by Kolmogorov and Arnold. One has to comprehend their work before trying to criticise the estimation of the random component in the CMB \art{wrandom}; and for that matter also~\art{GP}.

\end{document}